# *Antiferromagnetic MnNi tips for spin-polarized scanning probe microscopy*

P. R. Forrester, T. Bilgeri, F. Patthey, H. Brune[*], and F. D. Natterer[*]

Institute of Physics, École Polytechnique Fédérale de Lausanne, CH-1015 Lausanne, Switzerland

**Abstract**

Spin-polarized scanning tunneling microscopy (SP-STM) measures tunnel magnetoresistance (TMR) with atomic resolution. While various methods for achieving SP probes have been developed, each is limited with respect to fabrication, performance, and allowed operating conditions. In this study, we present the fabrication and use of SP-STM tips made from commercially available antiferromagnetic $Mn_{88}Ni_{12}$ foil. The tips are intrinsically SP, which is attractive for exploring magnetic phenomena in the zero field limit. The tip material is relatively ductile and straightforward to etch. We benchmark the conventional STM and spectroscopic performance of our tips and demonstrate their spin sensitivity by measuring the two-state switching of holmium single atom magnets on MgO/Ag(100).

## I. INTRODUCTION

Spin-polarized scanning tunneling microscopy (SP-STM) [1, 2] is an essential technique for investigating magnetism at the microscopic level, ranging from the study of magnetic skyrmions [3], single atoms and molecules [4, 5, 6] to the observation of Majorana modes [7, 8], spin filter measurements [9, 10], and the detection of electron spin resonance [11]. Moreover, SP-STM should facilitate further understanding of topological states in graphene systems [12, 13], topological insulators [14], and transition metal dichalcogenides [15]. Spin-polarized STM demands tips have a stable SP density of states (DOS) at the tunneling apex, in addition to being sufficiently sharp to achieve atomic resolution and spectroscopically stable over a large bias range for scanning tunneling spectroscopy (STS). An ideal SP-STM tip would also have a directionally adjustable magnetization component for in-plane and out-of-plane sensitivity. Although past efforts using bulk ferromagnets as tips and coating nonmagnetic tips with magnetic thin films yielded high spin contrast, they exhibited sizeable stray fields [16–20]. The functionalization of nonmagnetic tips with atoms or clusters of magnetic atoms also yields spin contrast, but is tedious, time consuming, and generally requires an external magnetic field to orient the paramagnetic tip-moment [16, 21, 22]. Tips fabricated from bulk antiferromagnetic materials, on the other hand, are intrinsically spin-polarized and tend to minimize the stray field. A large body of work focuses on chromium as a tip material [23–26], but the brittleness of Cr renders tip preparation difficult and motivates the search for alternative materials. Antiferromagnetic $Mn_{50}Ni_{50}$ also has been reported as a SP-STM probe material, but synthesis of this alloy requires specialized arc-melting [27]. The antiferromagnetic $Mn_{88}Ni_{12}$ alloy is an attractive alternative for SP-STM, as it contains a higher concentration of Mn, which is responsible for the spin sensitivity of the material [28–30] and is commercially available. We find that tips made from foils of $Mn_{88}Ni_{12}$ yield high spin contrast, are easy to etch electrochemically, are robust during handling, and perform equally well in topographic and spectroscopic modes as our previously used W and Pt/Ir tips.

## II. TIP FABRICATION

We fabricate $Mn_{88}Ni_{12}$ tips via a conventional loop method [31] in a homebuilt tip etching station shown in Fig. 1(a). Commercially available 0.25 mm thick $Mn_{88}Ni_{12}$ foil (Goodfellow) is cut into square rods via electrical discharge machining. We find this method yields superior rods compared to laser cutting. The latter typically results in irregular rod geometries that lead to undesired anisotropic etching and thus blunt tips, as suggested by Murphy *et al.* [27]. The cut rod is next lowered into a meniscus of 13% aqueous HCl, supported by an inert platinum ring. The rod is held at +10 V relative to the Pt ring, facilitating electrochemical etching of the region in contact with the solution. The gaseous products of the reaction frequently break the meniscus, requiring repeated rewetting of the loop. This does not appear to affect the tip quality as the meniscus re-forms around the etched area due to its high surface area. After about three minutes, the eroded area can no longer support the weight of the lower portion and breaks. This part lands on a bed of shaving cream that prevents damage to the atomically sharp candidate tip. The voltage is then immediately removed to slow the electrochemical reaction of the top part. Since both parts of the etched rods are tip candidates, we rinse them in deionized water, acetone, and isopropanol. We notice salt formation and further etching without the previous steps, possibly due to the reactivity of the etchant even without applied bias. We examine the candidate tips under a stereomicroscope. Typically, 20% of these tips are judged as sufficiently sharp to warrant further investigation with scanning electron microscopy (SEM). Figure 1(b) shows SEM micrographs of such a tip. The best tips are crimped onto a homemade tip holder, transferred into an ultra-high vacuum environment, and bombarded with a 1 kV beam of $Ar^+$ ions.

## III. SCANNING TUNNELING MICROSCOPY

### A. Scanning tunneling microscopy/spectroscopy

We evaluate the performance of the $Mn_{88}Ni_{12}$ tips on the model system of MgO/Ag(100) by reproducing topographic and spectroscopic characteristics detailed in previous studies [32–34]. Figure 2(a) shows a large-scale topographic STM image of MgO/Ag(100) measured with our $Mn_{88}Ni_{12}$ tip. We observe a smooth MgO layer and the nucleation of additional MgO islands, imaged as depressions due to their lower DOS up to the tunneling bias of 1 V. Our tip also resolves sharp atomic steps and isolated point defects over the large scan range in agreement with previous work [33]. The tips routinely achieve atomic resolution, as demonstrated in Fig. 2(b). The (1 x 1) oxygen sublattice of MgO is clearly reproduced as emphasized by the 2D FFT inset in Fig. 2(b).

We benchmark the spectroscopic performance of our $Mn_{88}Ni_{12}$ tips via STS. The *dI/dV* spectra recorded on MgO (red) and a Ti adatom (blue) are shown in Fig. 2(c). The essentially featureless MgO spectrum agrees with earlier results using non-SP tips [34] and demonstrates the smooth tip DOS. Our tip is thus sensitive to sample details and reproduces the large conductance steps of Ti at ~80 mV known from previous work [35]. In view of the well-behaved spectroscopy at low voltages, we examine the tip's performance at higher biases via field-emission resonance (FER) spectroscopy. Figure 2(d) shows a FER measurement of the MgO (red) and an MgO island (blue) in Fig. 2(a). We see an energetic downshift in the higher peak and an upshift in the lower peak with the addition of an MgO layer, in agreement with previous studies [33, 36].

**B. Spin-polarized scanning tunneling microscopy**

To test our $Mn_{88}Ni_{12}$ tips' spin-polarization, we dose holmium atoms (Inset of Fig. 3(a)) onto MgO, which are known to have a stable out-of-plane magnetization [5, 37]. They notably exhibit tunnel magnetoresistance through which the magnetization can be read non-destructively for tunnel voltages $V \leq 73$ mV [5]. Further, the Ho state may be switched for voltages above this threshold resulting in changes in TMR [5]. We exploit this phenomenon to prove the spin-sensitivity of our tip. The schematic in Fig. 3(a) illustrates the two magnetic states of Ho, Up and Down, with respect to the SP-DOS of the tip. When the majority spin occupation of the SP-DOS of the tip coincides with that of the Ho, the TMR is small (large current). Conversely, the TMR is large (small current) when the majority occupation of the SP-DOS of the tip and that of the Ho atom are not matched.

Figure 3(b) shows voltage dependent tunnel current-time traces, measured on MgO and Ho using our SP-tip. The current on the MgO substrate is constant, as expected. We also measure constant current on the Ho atom for a bias voltage of 60 mV. However, at higher voltages (80, 90, and 110 mV), we observe an accelerating two-state switching. In agreement with the aforementioned threshold voltage for magnetic switching [5], the current trace at 60 mV simply probes one of the two magnetic Ho states. Conversely, the two-state switching encountered for bias voltages above the switching threshold corresponds to the magnetization reversal of the Ho atom between its Up and Down states [5], manifested by a change in the TMR between the two configurations. The magnetic contrast is about 10% of the overall tunnel current. The switching rate increases monotonically with tunneling bias in agreement with previous measurements [5, 6]. The observed two-state switching of Ho, increase of the switching rate with increased bias, and stability of the Ho state below the threshold voltage unambiguously prove the magnetic sensitivity of our $Mn_{88}Ni_{12}$ tip in the out-of-plane direction.

Note that some microtip configurations yield no magnetic sensitivity to the switching of the holmium out-of-plane moment, but applying voltage pulses in the range of 1 to 10 V reliably yields SP microtips. We attribute this to two phenomena: (1) The magnetic moment of nickel atoms is quenched in the $Mn_{88}Ni_{12}$ bulk [38], thus a Ni terminated tip may not show SP; (2) the magnetic moment of a Mn terminated tip might be too oblique with respect to that of our Ho observer atom, thus leading to a reduced spin contrast in the out of plane direction that is insufficient to resolve changes in TMR. The orientation of the apex grain of the tip may be responsible for this. For the latter situation, we would obtain in-plane sensitivity instead, potentially allowing us to *in-situ* reconfigure a $Mn_{88}Ni_{12}$ tip for in-plane and out-of-plane measurements by voltage pulses. We find this tip curation procedure to be more efficient for $Mn_{88}Ni_{12}$ tips than for Cr tips, which inspired our preferential usage of the former.

The tip stray field is dominated by the apex atom, as inferred from a simple model calculation using the magnetic dipole-dipole interaction, lattice parameters, and magnetic moments of $Mn_{88}Ni_{12}$. In fact, a Hall measurement of an unetched rod of $Mn_{88}Ni_{12}$ shows no macroscopic field, as expected for a bulk antiferromagnet. We estimate a magnetic dipole stray field of ~14 mT for the measurement conditions in Fig. 3(b), assuming a tip moment of 2 $\mu_B$ at a distance of 6.5 Å between the centers of tip and sample atom (~2.5 Å away from point-contact) [28]. Due to the poor overlap between the 4*f* and *d* electron-wavefunctions, we do not consider exchange interactions [35].

## IV. SUMMARY AND OUTLOOK

We outlined the fabrication of SP-STM tips from commercially available $Mn_{88}Ni_{12}$ foil using one step HCl etching. We examined representative candidate tips with SEM, showing the necessary mesoscopic sharpness for STM studies. We atomically resolve MgO/Ag(100) and reproduce the known spectroscopic features of Ti. We proved the spin sensitivity of $Mn_{88}Ni_{12}$ tips by measuring the two state switching of Ho single atom magnets and described tip curation for achieving such spin contrast.

Tips made from $Mn_{88}Ni_{12}$ are promising for magnetic scanning probe experiments. Our work positively identifies $Mn_{88}Ni_{12}$ tips as a powerful, straightforward, and accessible alternative probe material for SPM that should enable the advanced study of magnetic phenomena in the absence of an applied magnetic field. These tips provide an avenue for examining magnetic phenomena at near zero effective field such as observing Majorana particles, discovering topologically protected states at interfaces, and reading and writing magnetic systems. Further, the relative ease with which $Mn_{88}Ni_{12}$ can be machined to small form factors is encouraging for more involved SPM techniques, such as magnetic force microscopy, where bulkier tips dramatically reduce the quality factor of the cantilever.

## V. EXPERIMENTAL DETAILS

All measurements are performed in a homebuilt STM [39] at a base pressure below $1 \times 10^{-10}$ mbar and at a temperature of 4.7 K. The Ag(100) surface is cleaned via alternating cycles of $Ar^+$ bombardment (~1 µA/cm$^2$) and annealing at 800 K. We grow MgO by dosing Mg from a Knudsen cell evaporator in an $O_2$ atmosphere of $1 \times 10^{-6}$ mbar onto the Ag(100) crystal held at 823 K at a rate of ~0.2 monolayers per minute. Titanium and Ho atoms are deposited from an e-beam evaporator onto the cooled ($T < 10$ K) sample in the STM position. All spectroscopic measurements are taken using a lock-in technique with $f_{mod} = 397$ Hz and $V_{mod} = 2$ mV$_{pp}$ except for the field emission resonances of MgO, which are measured with $f_{mod} = 1397$ Hz and $V_{mod} = 10$ mV$_{pp}$ with Z-feedback engaged. The current-time traces in Fig. 3(b) are measured with open Z-feedback at a nominal current setpoint of $I = 100$ pA and the voltages displayed in Fig. 3(b). We subtract a linear background from the current-time traces to account for drift in the tunnel junction. The averaging time for each current data point is 82 ms. All measurements are performed at zero applied field.

## ACKNOWLEDGMENTS

We thank the mechanical workshop of the EPFL Institute of Physics for expert technical assistance and the *Centre Interdisciplinaire de Microscopie Électronique* (CIME) for SEM characterization. P.R.F. appreciates support from the Fulbright U.S. Student Program. F.D.N., P.R.F., and T.B. thank the Swiss National Science Foundation for support under project numbers PZ00P2_167965 and 200020_176932.

**Figure Captions**

Fig. 1. (a) Photograph of tip etching station. A $Mn_{88}Ni_{12}$ rod is held at positive bias relative to a meniscus of HCl suspended by a platinum ring inducing the electrochemical erosion of the rod. When the eroded section can no longer support the weight of the lower segment of the rod, it breaks, and the lower segment lands on a mousse of shaving cream, preserving the sharp tip apex. (b) SEM micrographs of a representative $Mn_{88}Ni_{12}$ tip, imaged at a beam energy of 3 kV.

Fig. 2. (a) STM topographic image of MgO/Ag(100). Dark regions correspond to additional MgO islands ($V = 1$ V, $I = 100$ pA). (b) Atomically resolved STM image of MgO in (a) ($V = -20$ mV, $I = 10$ nA). Inset: 2D Fast Fourier Transform of (b). (c) $dI/dV$ point spectra of Ti (blue) and MgO (red) ($V = -100$ mV, $I = 200$ pA for Ti and $V = -100$ mV, $I = 500$ pA for MgO). Inset: STM image of a Ti atom on MgO ($V = -100$ mV, $I = 100$ pA). (d) $dI/dV$ spectra of MgO (red) and island (blue) ($V = 1$ V, $I = 100$ pA).

Fig. 3. (a) Tunneling magnetoresistance schematic showing larger current for matched tip and Ho majority and minority SP-DOS and smaller current for incommensurate SP-DOS, respectively. Inset: STM image of a holmium single atom magnet on MgO/Ag(100) ($V = -50$ mV, $I = 100$ pA). (b) Tunneling current-time trace measured on MgO (blue) at 150 mV and on Ho at increasing biases from 60 mV (yellow), to 110 mV (purple) at a nominal set-point current of 100 pA with Z-feedback open. The two-state switching of Ho occurs above a threshold voltage and is due to magnetization reversal of the Ho magnetic moment. Below the threshold, the orientation of the Ho moment is stable and no switching occurs.

Fig. 1.

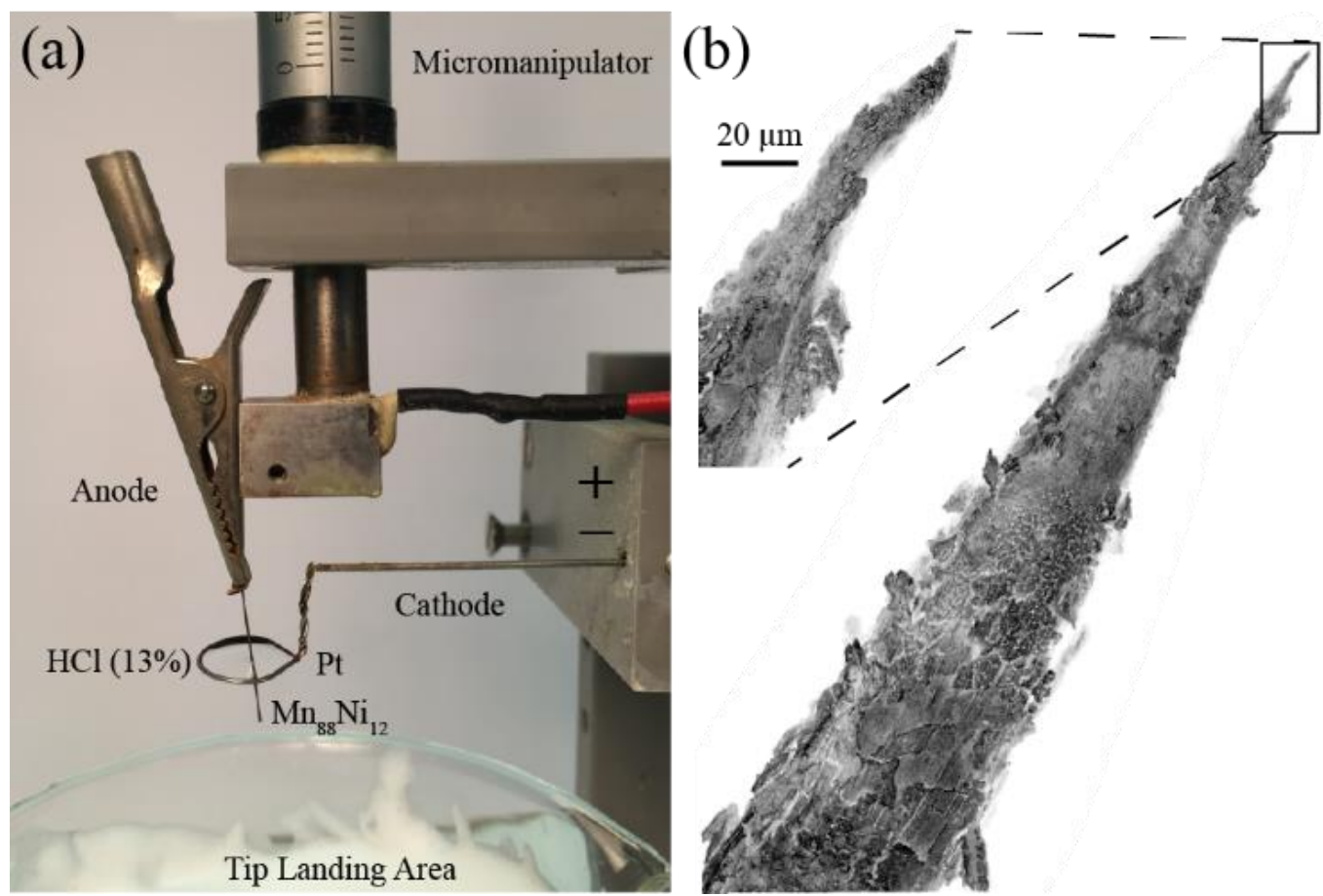

Fig. 2.

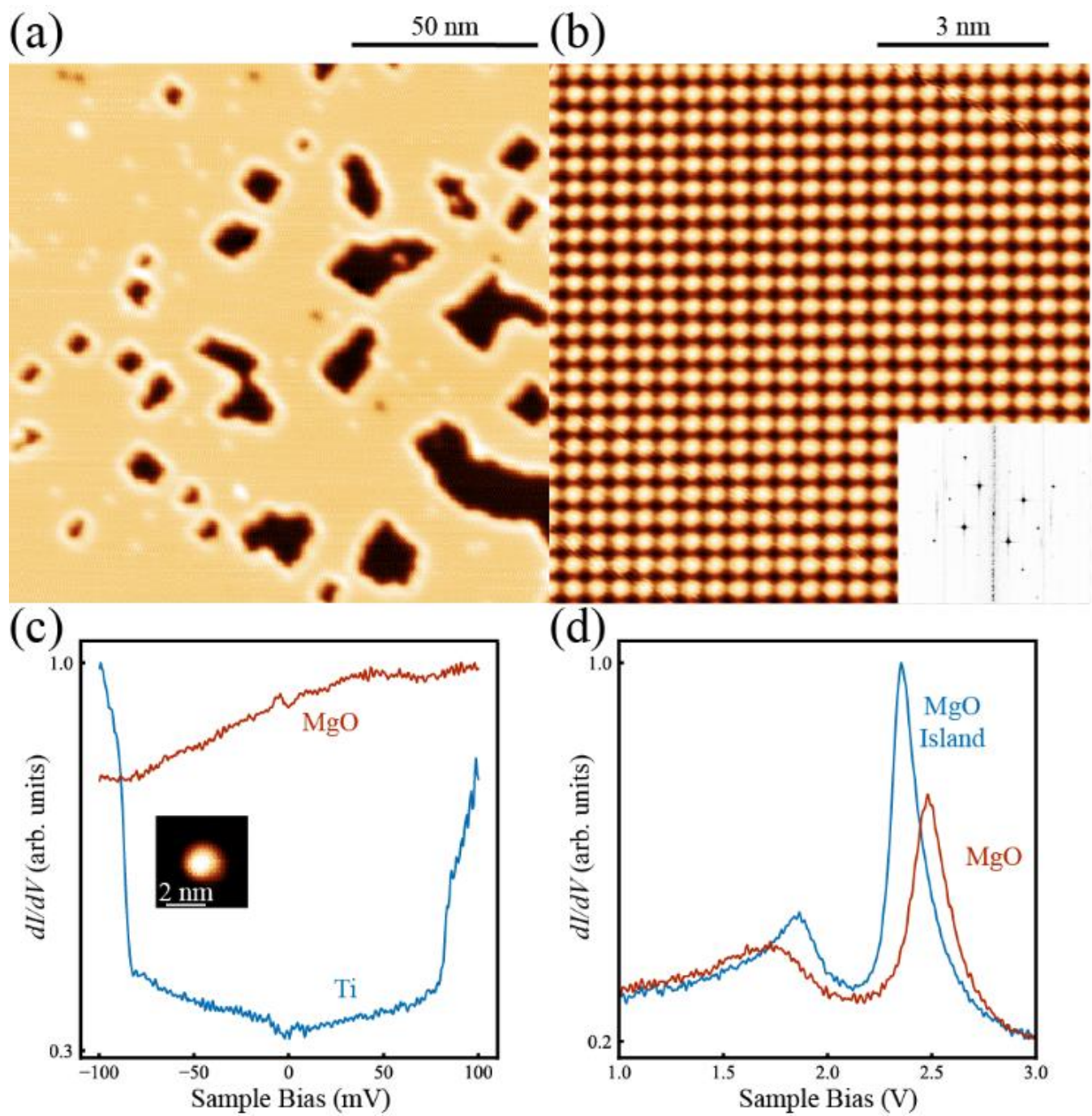

Fig. 3.

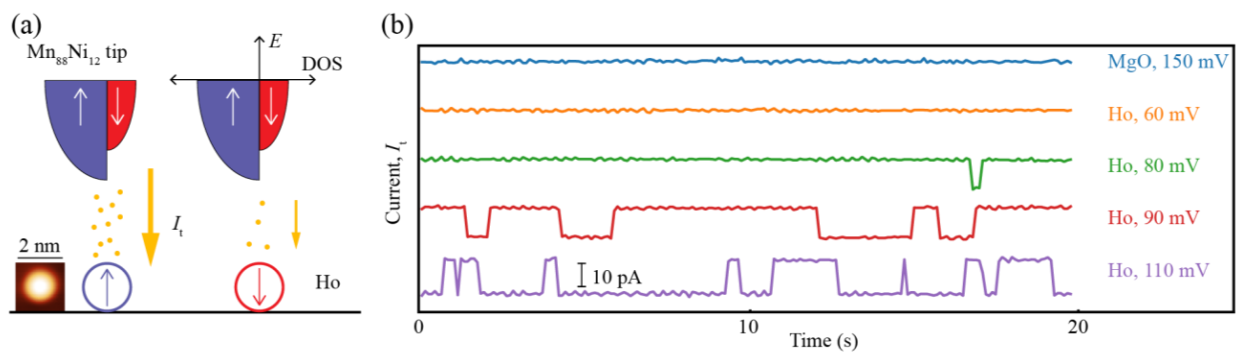